\documentclass[preprintnumbers,aps,superscriptaddress,showpacs,floatfix,nofootinbib]{revtex4-1}

\usepackage{hyperref}
\usepackage{color}
\usepackage{graphicx}	% Include figure filed
\graphicspath{%
  {./},%
}

\usepackage{amstext}
\usepackage{array}

\usepackage{xspace}	% Include xspace

\begin{document}

\preprint{ NT@UW-15-17}
\title{Testing Hydrodynamic Descriptions of p+p Collisions at $\sqrt{s}$=7 TeV}

\author{M.~Habich}
\affiliation{Department of Physics, 390 UCB, University of Colorado at Boulder, Boulder, CO, USA}
\author{G.A.~Miller}
\affiliation{Physics Department, University of Washington, Seattle, Washington 98195, USA}
\author{P.~Romatschke}
\affiliation{Department of Physics, 390 UCB, University of Colorado at Boulder, Boulder, CO, USA}
\affiliation{Center for Theory of Quantum Matter, University of Colorado, Boulder, Colorado 80309, USA}
\author{W.~Xiang}
\affiliation{Institute for Interdisciplinary Studies, Guizhou University of Finance and Economics, Guiyang 550025, China}
\affiliation{Department of Physics, 390 UCB, University of Colorado at Boulder, Boulder, CO, USA}

%\affiliation{University of Colorado at Boulder}
\date{\today}

\begin{abstract}
In high energy collisions of heavy-ions, experimental findings of collective flow are customarily associated with the presence of a thermalized medium expanding according to the laws of hydrodynamics. Recently, the ATLAS, CMS and ALICE experiments found signals of the same type and magnitude in  ultrarelativistic proton-proton collisions. In this study, the state-of-the-art hydrodynamic model SONIC is used to simulate the systems created in p+p collisions. By varying the size of the second-order transport coefficients, the range of applicability of hydrodynamics itself to the systems created in p+p collisions is quantified. It is found that hydrodynamics can give quantitatively reliable results for the particle spectra and the elliptic momentum anisotropy coefficient $v_2$. Using a simple geometric model of the proton based on the elastic form factor leads to results of similar type and magnitude to those found in experiment when allowing for a small bulk viscosity coefficient.
\end{abstract}

\maketitle

%\subsection{Introduction}

The experimental heavy-ion program at the Relativistic Heavy Ion Collider (RHIC) and the Large Hadron Collider (LHC) has provided strong evidence for the creation of an equilibrated state of matter in ultrarelativistic collisions of heavy-ions such as gold or lead \cite{Adams:2005dq,Adcox:2004mh,Back:2004je,Arsene:2004fa,Aamodt:2010pa,Aad:2014vba,Chatrchyan:2011sx}. Comparing the wealth of experimental data available over a large range of collision energies to theoretical model calculations, the current consensus in the field is that the matter created in ultrarelativistic heavy-ion collisions behaves like an almost ideal fluid with very low shear viscosity over entropy ratio \cite{Huovinen:2001cy,Teaney:2003kp,Hirano:2005xf,Luzum:2008cw,Schenke:2010rr,Heinz:2013th}. This form of matter has been dubbed the 'quark-gluon plasma'. 

Only a few years ago, there was a similar consensus in the field that the systems created in proton-nucleus collisions (or d+Au collisions in the case of RHIC) did not equilibrate to form a quark gluon plasma because these systems were too small, too short-lived, and contained too few particles to behave collectively. In fact, experimental data from these light-on-heavy ion collisions was regarded as a reference system in which the quark-gluon plasma component was 'known' to be absent. Similarly, the notion that quark-gluon plasmas could be formed in high energy proton-proton collisions was mostly regarded as preposterous: how could a system consisting of a handful of particles behave as a fluid? 

The consensus in the field was severely challenged, if not shattered, when experimental data for anisotropic collective flow in p+Pb, p+Au, d+Au, $^3$He+Au, and most recently in proton-proton collisions became available \cite{Abelev:2012ola,Aad:2012gla,Adare:2013piz,Adare:2015ctn,Aad:2015gqa}. In all of these small systems, the experimental signals turned out to be similar in type and magnitude to those found in heavy-ion collisions. Furthermore, the measurements could again be well described (and in some cases predicted) by theoretical hydrodynamic model calculations \cite{Bozek:2011if,Nagle:2013lja,Schenke:2014zha,Kozlov:2014fqa,Romatschke:2015gxa}, such as the SONIC model \cite{Habich:2014jna}. 

The experimental finding of a large elliptic flow coefficient $v_2$ in high energy proton-proton collisions is particularly intriguing, because a large $v_2$ coefficient is typically indicative of a hydrodynamic phase in the system evolution \cite{Romatschke:2015dha}. Is it at all possible for hydrodynamics to quantitatively describe the real-time evolution of system with a linear dimension of less than $1$ fm and an average of five to six particles per unit rapidity? What constraints would result on QCD transport coefficients such as shear and bulk viscosity? These questions provide the motivation for performing a hydrodynamic study of high-energy proton-proton collisions. 

One of the key differences of the present study with respect to most previous hydrodynamic studies of proton-proton collisions such as those in Refs.~\cite{Prasad:2009bx,Bozek:2009dt, Ortona:2009yc, Werner:2010ss} is the inclusion of both shear and bulk viscous effects in the hydrodynamic evolution. (Note that shear viscous effects were already included in Ref.~\cite{Luzum:2009sb}, which will be discussed below in more detail). Another perhaps novel aspect of the present study is that 'typical' proton-proton collisions (as opposed to high-multiplicity events such as those studied in Ref.~\cite{CasalderreySolana:2009uk}) will be discussed. Finally, the main emphasis of the present study will be a quantitative test of applicability of hydrodynamics to small systems, which has never been attempted before.

\section{Methodology}

In the present study, we use the hydrodynamic model SONIC \cite{Habich:2014jna} to simulate the matter created in proton-proton collisions. SONIC simulates the dynamics in the plane transverse to the beam axis using causal relativistic hydrodynamics in the presence of shear and bulk viscosity, followed by the hadron cascade afterburner B3D \cite{Novak:2013bqa} in the hadronic phase for temperatures $T<0.17$ GeV, while assuming boost-invariance in the longitudinal direction (see Ref.~\cite{Habich:2014jna} for a detailed discussion of SONIC's components). It should be noted that while SONIC implements shear viscous effects when switching from hydrodynamics to the hadron cascade simulation \cite{Pratt:2010jt}, the consistent implementation of bulk viscous effects on particle spectra is currently poorly understood \cite{Monnai:2009ad}. For this reason, bulk viscous contributions to the initial particle spectra in the hadron cascade are not included in the present description. This is different from other works in the literature (e.g. \cite{Bozek:2009dw,Ryu:2015vwa}) which use a form of the bulk viscous corrections based on a quasi-particle model \cite{Sasaki:2008fg}.

SONIC is known to successfully describe experimental data for p+Pb and d+Au collisions at $\sqrt{s}=5.02$ TeV and $\sqrt{s}=0.2$ TeV collision energies, respectively, and has been used to make accurate predictions for $v_2,v_3$ for $^3{\rm He}$+Au collisions and p+Au collisions at $\sqrt{s}=200$ GeV \cite{Nagle:2013lja,Romatschke:2015gxa}. 
In order to simulate proton-proton collisions, a model for the hydrodynamic initial conditions, such as the energy density distribution in the transverse plane, is needed. These initial conditions are poorly constrained from first-principles calculations, so a basic model built on the proton form factor was used, which are described below. Besides the initial conditions, the hydrodynamic evolution in SONIC requires specification of the simulated ratios of shear viscosity and bulk viscosity to entropy density, $\frac{\eta}{s},\frac{\zeta}{s}$, respectively. In the following, both of these ratios were taken to be constant in temperature for simplicity. Finally, SONIC requires specification of second order transport coefficients, such as the shear and bulk relaxation times $\tau_\pi,\tau_\Pi$, respectively (cf. Ref.~\cite{Romatschke:2009kr}). For simplicity, we have set $\tau_\pi=\tau_\Pi$ (cf. Ref.~\cite{Kanitscheider:2009as}).

The value of these relaxation times controls the size of second-order gradient terms in the hydrodynamic expansion. Varying the relaxation times thus allows one to quantify the importance of second-order gradient terms in final results, and thus provides a measure of the quantitative reliability of the hydrodynamic gradient expansion. The ``conventional'' criterion for the applicability of hydrodynamics states that the mean free path $\lambda$ needs to be much smaller than the system size $L$. The ratio $\frac{\lambda}{L}$ is referred to as Knudsen number, and the conventional criterion quantifies the size of first-order gradient corrections (viscous effects) to ideal hydrodynamics. In recent years there has been mounting evidence from exact solutions of far-from equilibrium quantum field theories that (second-order) hydrodynamics quantitatively applies in cases where first-order (viscous) corrections to ideal hydrodynamics are large (order unity, cf. Refs.~\cite{Heller:2011ju,Wu:2011yd,vanderSchee:2012qj,Chesler:2015bba}). Thus it may be that the ``conventional'' Knudsen number criterion considerably underestimates the applicability of hydrodynamics. Instead, it has been suggested that the true criterion for the applicability of hydrodynamics is set by the location of the first non-hydrodynamic singularity in the complex frequency plane \cite{Heller:2013fn}. In second-order hydrodynamics, the location of this pole is controlled by the value(s) of the relaxation time. Hence it is plausible that varying the relaxation time $\tau_\pi$ allows a modern, realistic, quantitative and easily implementable test for the applicability of hydrodynamics. This is consistent with the notion of large first, but small second-order hydrodynamic corrections.

It is well known that for fixed shear-viscosity over entropy ratio, the value of $\tau_\pi$ varies very little (only by about a factor of two) when the interaction strength in a  quantum field theory is changed from zero to infinity \cite{Baier:2007ix,York:2008rr}. With this result in mind, we choose to quantify the applicability of hydrodynamics by varying the relaxation times by 50 percent around a fiducial value of $\tau_\pi=6 \frac{\eta}{s T}$. If the resulting variations in the final results are large, then hydrodynamics does not apply. Conversely, if the variations turn out to be small, then this provides evidence that hydrodynamics can give a quantitatively reliable description of the system.

{\bf Basic model for the proton:} We consider the initial transverse energy density distribution $\varepsilon$ to be given by
\begin{equation}
\label{eq:optg}
\varepsilon(x,y,\tau_0)=\kappa(\tau_0) T_1\left(x+\frac{b}{2},y\right)T_2\left(x-\frac{b}{2},y\right)\,,
\end{equation}
where ${\bf x_\perp}=(x,y)$ are the coordinates in the transverse plane, $\tau_0$ is the initialization time of hydrodynamics, $b$ is the impact parameter of the collision, $\kappa(\tau_0)$ is an overall normalization that is fixed by the experimental multiplicity in minimum-bias collisions, and $T_{1,2}$ is the transverse charge density distribution of proton $1$ and $2$, respectively. The expert reader will recognize Eq.~(\ref{eq:optg}) as an optical-Glauber model for protons, where it should be pointed out that for protons the binary collision scaling coincides with the number of participants scaling because $A=1$. Indeed, in the basic initial condition model (referred to as 'RND' for 'round' in the following), we take $T(x,y)$ to be given by the Fourier-transform of the proton form factor $F(Q^2)$,
\begin{equation}
\label{eq:Trnd}
T_{\rm RND}({\bf x_\perp})=\int \frac{d^2 q}{(2 \pi)^2} e^{-i {\bf q}\cdot{\bf x_\perp}} F(Q^2={\bf q}^2)\,,
\end{equation}
where we take the parametrization of the form factor from Ref.~\cite{Venkat:2010by}. In the RND model, the proton is always round, and initial conditions for $\varepsilon$ are generated by Monte-Carlo sampling of impact parameters $b \in [0,b_{\rm max}]$, where the upper limit $b_{\rm max}=1.6$ fm corresponds to approximately twice the proton radius.

In a variation of the 'RND' model for initial conditions, referred to as 'FLC' for 'fluctuating' in the following, spin fluctuations of the proton are considered. Using the model from Ref.~\cite{Miller:2003sa}, the overlap function is defined as
\begin{eqnarray}
  T_{\rm FLC}({\bf x_\perp})&=&\int_{-\infty}^\infty dz\left[\frac{\rho_U(r)\left(1+\hat{n}\cdot \hat{s}\right)}{2 N} \right.\nonumber\\
&&\left.+\frac{\rho_L(r)\left(1+2 \hat{r}\cdot\hat{s}\hat{r}\cdot\hat{n}-\hat{n}\cdot\hat{s}\right)}{2 N}\right]\,,
\end{eqnarray}
where ${\bf r}=(x,y,z)$, $r=|{\bf r}|$ and $N=4 \pi \int_0^\infty dr r^2\left[\rho_U(r)\frac{1+\hat{n}\cdot \hat{s}}{2}+\rho_L(r)\frac{3-\hat{n}\cdot\hat{s}}{6}\right]$ is a normalization to ensure that protons have electric charge of unity for arbitrary unit vectors $\hat s,\hat n$. In the FLC model, the proton's shape may fluctuate event-by-event, and initial conditions for $\varepsilon$ are generated by Monte-Carlo sampling of the two unit vectors $\hat s,\hat n$ as well as the impact parameter of the collision $b \in [0,b_{\rm max}]$.

\section{Results}

\begin{figure*}[t]
\centering
\includegraphics[width=0.45\textwidth]{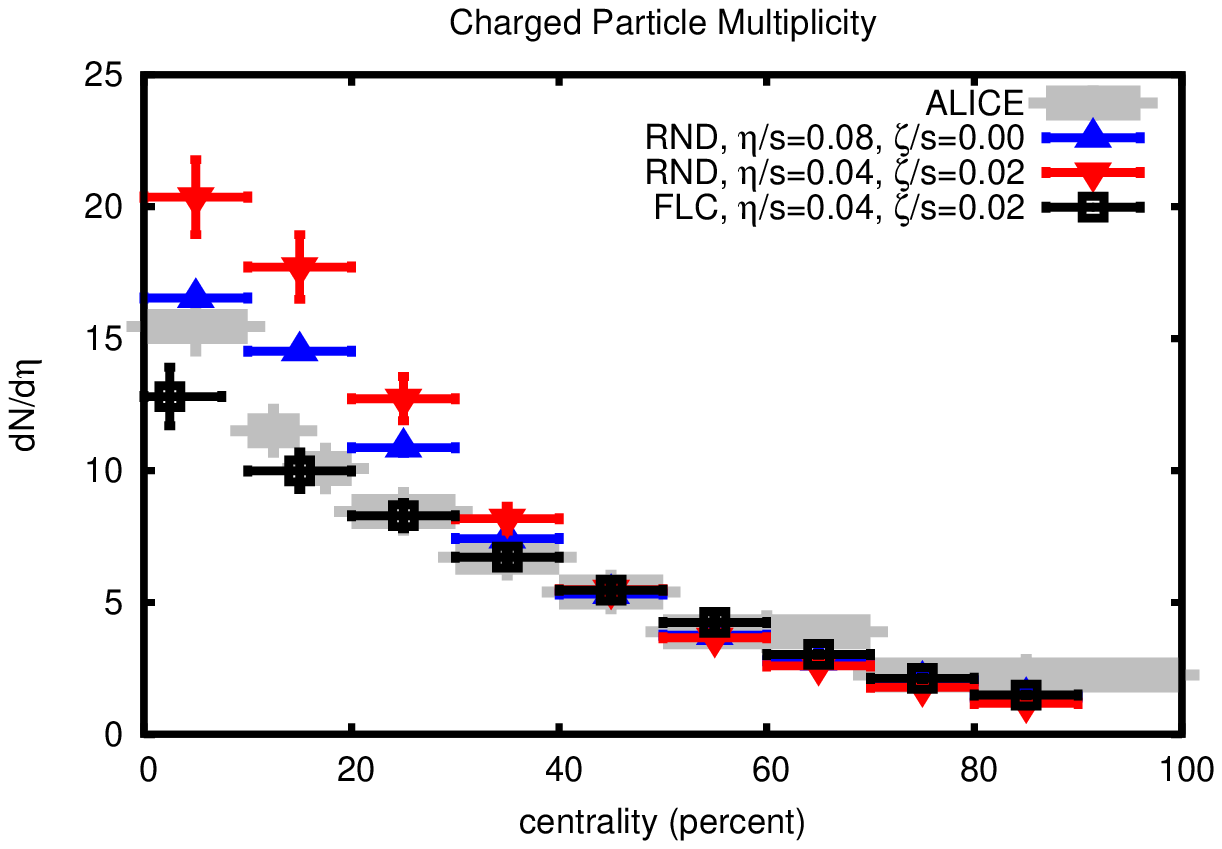}\hfill
\includegraphics[width=0.45\textwidth]{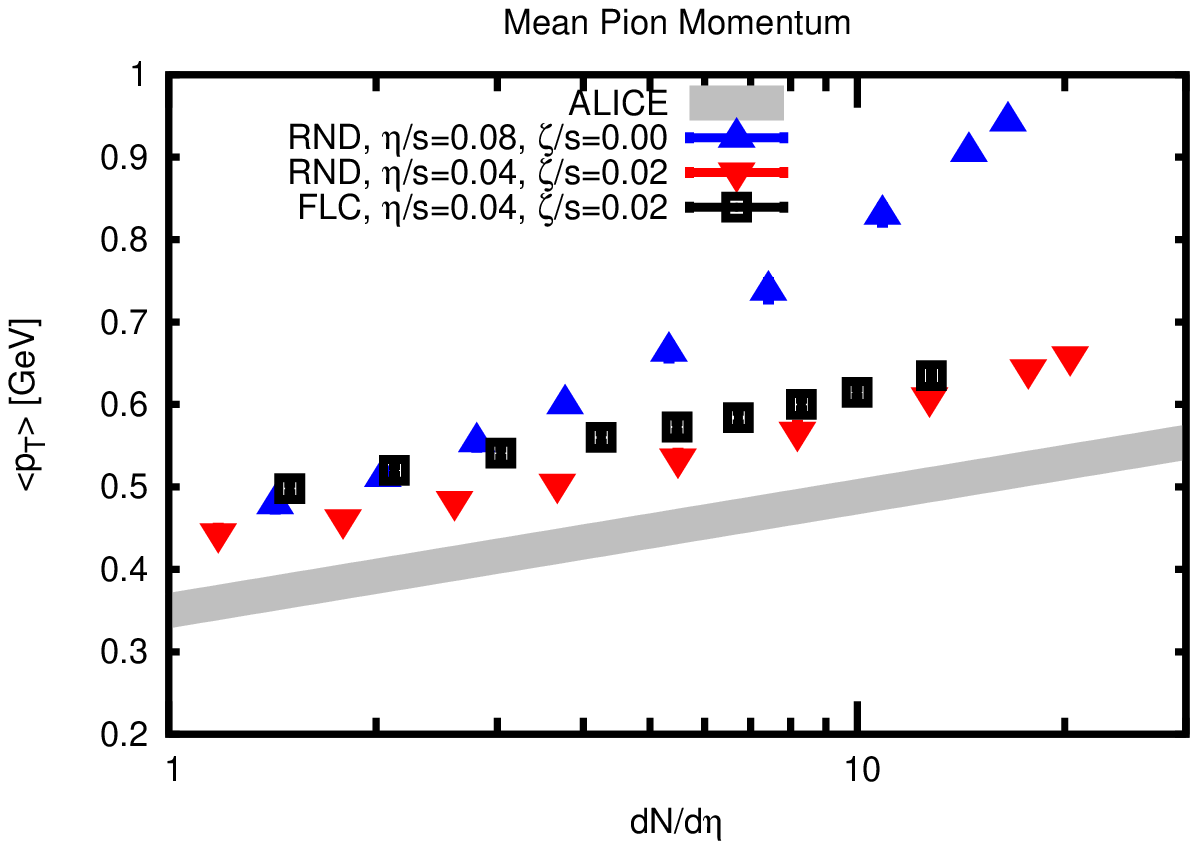}
\caption{Unidentified charged hadron multiplicity (left) and pion mean transverse momentum (right) for p+p collisions at $\sqrt{s}=7$ TeV. Shown are experimental results from ALICE (cf. \cite{LivioTalk}) and SONIC simulations for proton models based on the proton form factor. The error bars for the SONIC simulations include systematic uncertainties for the applicability of hydrodynamics obtained from varying second-order transport coefficients; as can be seen, those error bars are significant for neither the multiplicity nor the pion $<p_T>$, thus indicating robust applicability of hydrodynamics for these quantities. Note that the 'RND' model has been run with different shear and bulk viscosities. While the effect of changing the shear viscosity on the multiplicity and transverse momentum is minor (not shown), even a very small bulk viscosity has a large effect on the final pion transverse momentum. 
\label{fig:first}}
\end{figure*}

Using the basic model of the proton described in the previous section, the hydrodynamic plus cascade model SONIC was initialized at $\tau_0=0.25$ fm/c and results for particle spectra and momentum anisotropies were obtained that can be directly compared to experimental measurements (cf.~\cite{Habich:2014jna}).
In Fig.~\ref{fig:first}, results for the multiplicity of unidentified charged hadrons\footnote{In the simulation, $\frac{dN}{dY}$ is reduced by ten percent to obtain the experimentally determined pseudo-rapidity distribution $\frac{dN}{d\eta}$.} and mean pion transverse momentum are shown for proton-proton collisions at $\sqrt{s}=7$ TeV. The multiplicity in the 40--50 percent centrality class obtained by ALICE \cite{LivioTalk} was used to set the overall constant $\kappa$ in the SONIC simulations. The error bars shown for the SONIC results include the systematic uncertainties for the applicability of hydrodynamics obtained from varying second-order transport coefficients, as described above. From Fig.~\ref{fig:first} it becomes apparent that systematic uncertainties of hydrodynamics for the particle multiplicity and mean transverse momentum are small, providing evidence that a hydrodynamic description of these quantities is feasible for proton-proton collisions. The centrality dependence of multiplicity in SONIC is broadly consistent with the experimental measurements from ALICE, with a level of disagreement that can be expected given the simplicity of the initial conditions used. Considering the mean transverse pion momentum, Fig.~\ref{fig:first} indicates that SONIC results are extremely sensitive to the presence of bulk viscosity, as is apparent from comparing the 'RND' model results for $\frac{\zeta}{s}=0$ and $\frac{\zeta}{s}=0.02$. This effect originates from the modification of the fluid flow from bulk viscosity, and thus is expected to be a robust feature irrespective of the hadronization prescription used (see also the discussion in Appendix \ref{sec:appa}). For the proton models used, a minimum non-zero value of $\frac{\zeta}{s}$ was needed to bring any of the  theory calculation close to the experimental data from the ALICE experiment \cite{LivioTalk} for pion mean transverse momentum. Because of the crudeness of the proton model, no effort has been made to tune transport coefficient in order to match the experimental data.

\begin{figure*}[t]
\centering
\includegraphics[width=0.45\textwidth]{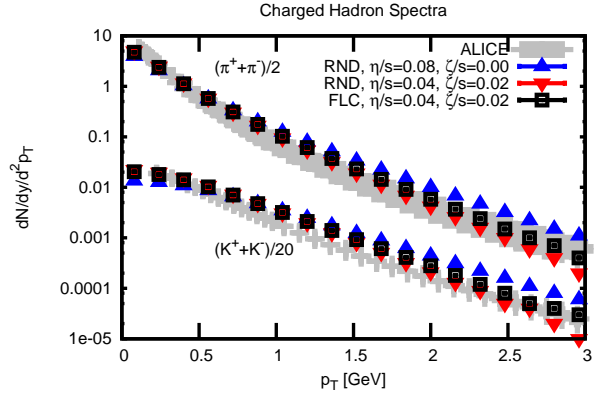}\hfill
\caption{Pion and kaon spectra for the 40-50 percent centrality class compared to measured minimum bias spectra for $\sqrt{s}=7$ TeV from the ALICE experiment \cite{Adam:2015qaa}. The error bars for the SONIC simulations include systematic uncertainties for the applicability of hydrodynamics obtained from varying second-order transport coefficients; these error bars are smaller than the symbol size for particle spectra, thus indicating robust applicability of hydrodynamics for this quantities. Note that the 'RND' model has been run with different shear and bulk viscosities, indicating the sensitivity of particle spectra to a small bulk viscosity coefficient.
\label{fig:spec}}
\end{figure*}

Comparisons of identified particle spectra for mid-central collisions to minimum-bias experimental data are shown in Fig.\ref{fig:spec}. Again, one observes reasonable overall agreement between simulations and experiment except for the case when bulk viscosity was set to zero.

The qualitative effect of bulk viscosity reducing the mean particle momenta was observed before in heavy-ion collisions, e.g. in \cite{Monnai:2009ad,Ryu:2015vwa}. However, the effect of including the bulk viscosity in proton-proton collisions is much more pronounced than in heavy-ion collisions. Specifically, we find a factor two decrease in pion momentum originating from a bulk viscosity coefficient of $\frac{\zeta}{s}=0.02$, while Ref.~\cite{Ryu:2015vwa} found approximately 25 percent reduction for a bulk viscosity coefficient peaking at $\frac{\zeta}{s}=0.3$ (note that such high values would likely cause cavitation in the fluid \cite{Rajagopal:2009yw,Habich:2014tpa,Sanches:2015vra}).

\begin{figure*}[t]
\centering
\includegraphics[width=0.45\textwidth]{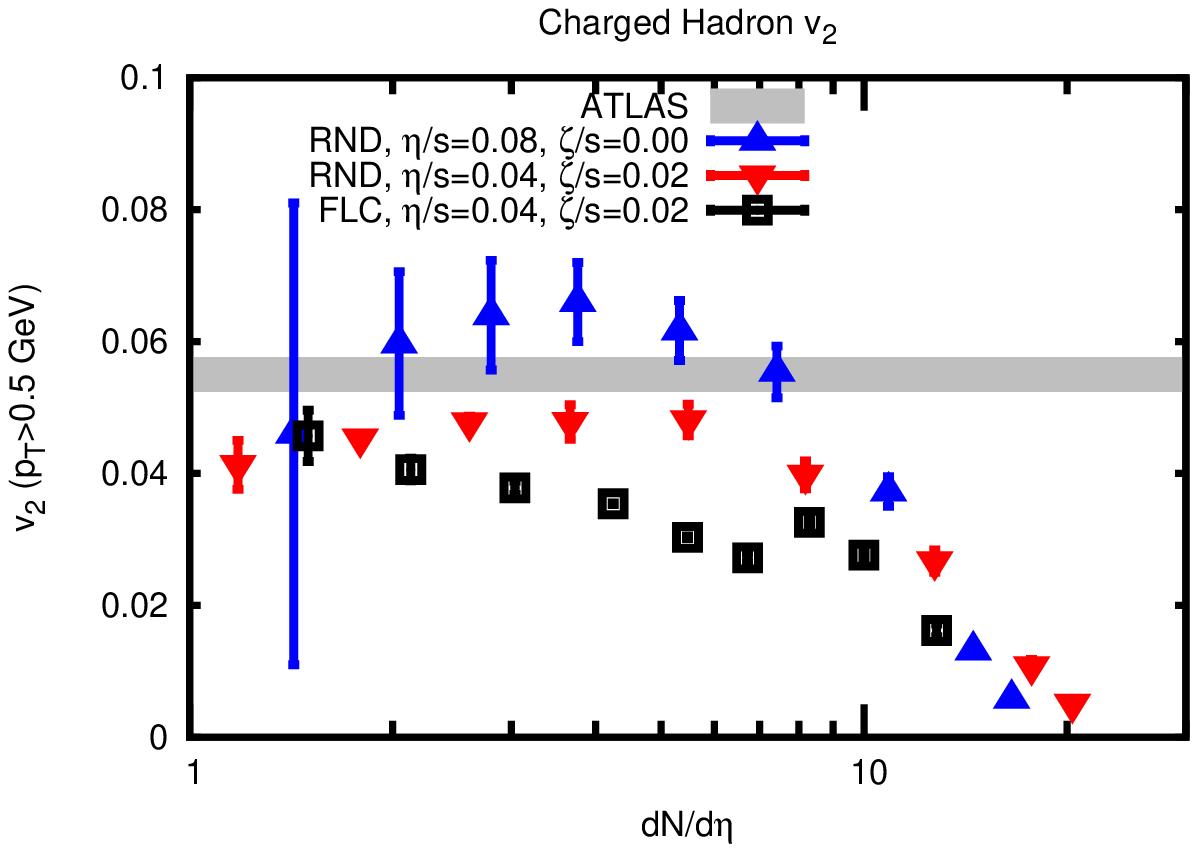}\hfill
\includegraphics[width=0.45\textwidth]{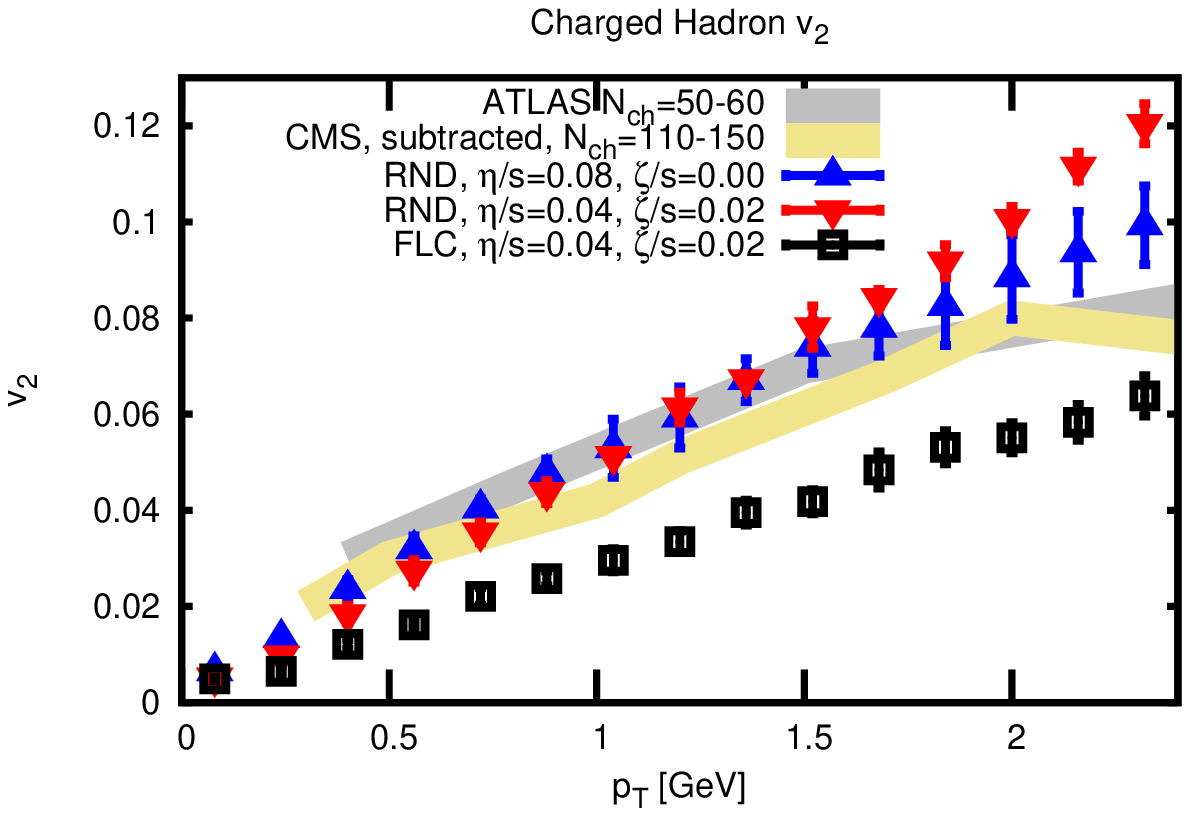}
\caption{Left: Integrated momentum anisotropy $v_2$ for unidentified charged hadrons with $p_T>0.5$ GeV in proton-proton collisions. Shown are the range of experimental results from ATLAS (cf. \cite{Aad:2015gqa}) for $\sqrt{s}=2.76,13$ TeV and SONIC simulations for $\sqrt{s}=7$ TeV. The error bars for the SONIC simulations include systematic uncertainties for the applicability of hydrodynamics obtained from varying second-order transport coefficients. Right: Unintegrated momentum anisotropy for unidentified charged hadrons for the 40-50 percent centrality class compared to experimental results from ATLAS \cite{Aad:2015gqa} with $N_{\rm ch}=50-60$ and for the 0.5-4 percent centrality class ($N_{\rm ch}=110-150$) from CMS \cite{CMS:2015kua}. We expect the $v_2(p_T)$ result from the  40-50 percent centrality class in our simple proton models to be most representative of the experimental results for all centralities, including central collisions.
\label{fig:second}}
\end{figure*}

In Fig.~\ref{fig:second}, the momentum anisotropy coefficient $v_2$ for unidentified charged particles with $p_T>0.5$ GeV from SONIC, including the estimated systematic uncertainty from the hydrodynamic gradient expansion is shown. (Note that $v_2$ is considerably smaller when a smaller $p_T$ cut is used, cf. Ref.~\cite{Luzum:2010ag}). As outlined in the methodology section above, a large systematic uncertainty compared to the mean value indicates that hydrodynamic is very sensitive to the detailed treatment of higher order gradient terms and/or non-hydrodynamic degrees of freedom. Thus a large uncertainty signals the breakdown of hydrodynamics. There are no established criteria in the literature for what constitutes an unacceptably large uncertainty, so in the following we declare a breakdown of hydrodynamics to occur if the ratio of uncertainty to mean value exceeds 50 percent. In the case of the $v_2$ values shown in the left-hand panel of Fig.~\ref{fig:second}, this threshold is reached  for $\frac{dN}{d\eta}\lesssim 2$ and $\frac{\eta}{s}=0.08$, indicating that the hydrodynamic description of $v_2$ has broken down in this case.

On the other hand, while the systematic uncertainty originating from higher order gradient terms is sizable, it seems that hydrodynamics nevertheless is still applicable to describing $v_2$ in proton-proton collisions for $\frac{dN}{d\eta}\gtrsim 2$ when $\frac{\eta}{s}\leq 0.08$. Since this finding disagrees with an earlier prediction by one of us in Ref.~\cite{Luzum:2009sb}, this point deserves further clarification. Unlike the earlier study in Ref.~\cite{Luzum:2009sb}, the present study does not use hydrodynamics for temperatures below the QCD phase transition, but instead employs a hadronic cascade simulation, thus increasing overall reliability of the model.  

As can be seen by e.g. comparing the results for $\frac{\eta}{s}=0.08$ and $\frac{\eta}{s}=0.04$ in Fig.~\ref{fig:second}, the hydrodynamic systematic uncertainties decrease when lowering $\frac{\eta}{s}$. This is a trivial consequence of the fact that uncertainties are calculated by varying $\tau_\pi$ and $\tau_\pi\propto \frac{\eta}{s}$, so decreasing $\frac{\eta}{s}$ also decreases the extent of the variation. In the ideal hydrodynamic limit when $\frac{\eta}{s}\rightarrow 0$, second-order hydrodynamics no longer depends on the relaxation time nor does it possess a non-hydrodynamic pole, so an effective ideal hydrodynamic description never breaks down. This somewhat counter-intuitive finding can be justified physically by noting that in the ideal hydrodynamic limit, the mean free path $\lambda$ tends to zero, so that even for very small system sizes $L$ (or strong gradients) one always has $\frac{\lambda}{L}\rightarrow 0$. There are strong indications to support the notion that a lower bound on $\frac{\eta}{s}$ exist, effectively prohibiting to ever reach the ideal hydrodynamic limit in practice. However, this information is not part of a hydrodynamic description or the calculation of systematic uncertainties in this framework.

Also shown in Fig.~\ref{fig:second} is the range of experimental results for $v_2$ as measured by the ATLAS experiment \cite{Aad:2015gqa} for p+p collisions at $\sqrt{s}=2.76$ TeV and $\sqrt{s}=13$ TeV for $N_{\rm ch}=50-60$, which roughly corresponds to the 0.5--4\% centrality class (cf.~\cite{CMS:2015kua}). The SONIC model simulation results include no (RND) or only limited (FLC) event-by-event fluctuations, thereby invalidating the model results for the most central collisions ($\frac{dN}{d\eta}\gtrsim 10$) and the most peripheral collisions ($\frac{dN}{d\eta}\lesssim 1$). For mid-central collisions, however, the 'RND' and 'FLC' model are broadly consistent with the magnitude of the measured $v_2$ coefficient by the ATLAS experiment. This finding is corroborated by the second panel in Fig.~\ref{fig:second}, where the momentum dependence of the $v_2$ coefficient for mid-central collisions (40-50 percent centrality class) is compared to experimental data for more central collisions from the ATLAS and CMS experiments.

SONIC simulation results for $v_2$ are sensitive to both shear and bulk viscosity coefficients, and no attempt has been made to tune the value of those coefficients in order to match the experimental data in view of the crudeness of the initial condition model. Rather, one observes that with 'typical' values for $\frac{\eta}{s},\frac{\zeta}{s}$ the SONIC model predicts a $v_2$ response that is of comparable to that measured by experiment.

\section{Conclusions}

The hydrodynamic model SONIC was used to study proton-proton collisions at $\sqrt{s}=7$ TeV by employing a simple parametrization of proton based on the elastic form factor. By varying the size of the second-order transport coefficients, the applicability of hydrodynamics itself to the systems created in p+p collisions could be quantified. It was found that a hydrodynamic description of the momentum anisotropy coefficient $v_2$ is breaking down for $\frac{dN}{d\eta}\lesssim 2$ when $\frac{\eta}{s}\geq 0.08$. Conversely, it was found that hydrodynamics 
can give quantitatively reliable results for the particle spectra and the elliptic momentum anisotropy coefficient $v_2$ when $\frac{dN}{d\eta}\gtrsim 2$. While it is somewhat surprising that hydrodynamics applies even for such low multiplicities, this finding is qualitatively in line with recent results for proton-nucleus collisions in Ref.~\cite{Romatschke:2015gxa}. In Ref.~\cite{Romatschke:2015gxa} it was found that a hydrodynamic description of $v_2$ was found to be reliable whereas hydrodynamics would break down sequentially starting from the higher order momentum anisotropies (first $v_5$, then $v_4$, etc). The finding that hydrodynamics can be applied to proton-proton collisions is also consistent with recent results from gauge/gravity duality simulations in Ref.~\cite{Chesler:2015bba}. This surprising applicability of hydrodynamics to small systems becomes somewhat less mysterious if one abandons the traditional idea of a handful of quarks and gluons forming a fluid in favor of delocalized and strongly interacting fields forming a plasma. Since hydrodynamics can be derived from a gradient expansion of quantum field theory without ever employing the concept of quasi-particles \cite{Baier:2007ix,Bhattacharyya:2008jc}, it is perfectly reasonable to expect a tiny droplet of deconfined and strongly interacting QCD matter to behave hydrodynamically, even if this droplet will eventually hadronize into only a handful of hadrons. In principle, this notion could even offer a new interpretation of the apparently thermalized particle spectra seen to $e^+$+$e^-$ collisions.

In the context of a hydrodynamic description, the present study provided evidence that final particle mean transverse momenta in p+p collisions are strongly sensitive to the bulk viscosity coefficient. A non-vanishing minimum value of $\frac{\zeta}{s}$ was required to match experimental measurements of mean transverse momentum. This could indicate a possible experimental path to determining the bulk viscosity coefficient in QCD. Finally, it was found that typical elliptic momentum anisotropy coefficients $v_2$ obtained in the hydrodynamic model are of the same magnitude as those measured by experiment. 

Clearly, many aspect of the present hydrodynamic study could and should be improved when aiming at a detailed description of experimental data in the future, such as the inclusion of more realistic event-by-event fluctuations for the proton shape, or pre-equilibrium flow. However, we do not expect these future improvements of the treatment of initial conditions to affect the applicability of hydrodynamics.

To conclude, our study provides evidence that the experimental results obtained in high energy proton-proton collisions can be understood both qualitatively and quantitatively in terms of a hydrodynamic model similar to that used in heavy-ion collisions. While the present hydrodynamic model does not describe details of the experimental measurements, it is likely that more sophisticated parametrizations of the proton could bring the same level of agreement to proton-proton collisions as is now routinely seen in heavy-ion collisions. This implies that an interpretation of the formation of a quark-gluon plasma in proton-proton collisions is consistent with the experimental data, yet does not imply that it is the only such consistent interpretation. Future work is needed to improve our qualitative and quantitative understanding of these fascinating system that link the fields of high energy and nuclear physics.

\begin{acknowledgments}
%\paragraph{\bf Acknowledgements:}
 
The work of G. A. M.    was supported by the U. S. Department of Energy Office of Science, Office of Nuclear Physics under Award Number DE-FG02-97ER-41014. The work of M.H. and P.R. was supported by the U.S. Department of Energy, DOE award No. DE-SC0008132. W.X. was supported by National Natural Science Foundation of China No.11305040 and thanks the Department of Physics The University of Colorado at Boulder for the hospitality when this work was completed. We would like to thank M.~Floris for fruitful discussions.

\end{acknowledgments}

\begin{appendix}

\section{Bulk Viscous Effects on Hydrodynamic Flow}
\label{sec:appa}

\begin{figure*}[b]
\centering
\includegraphics[width=0.45\textwidth]{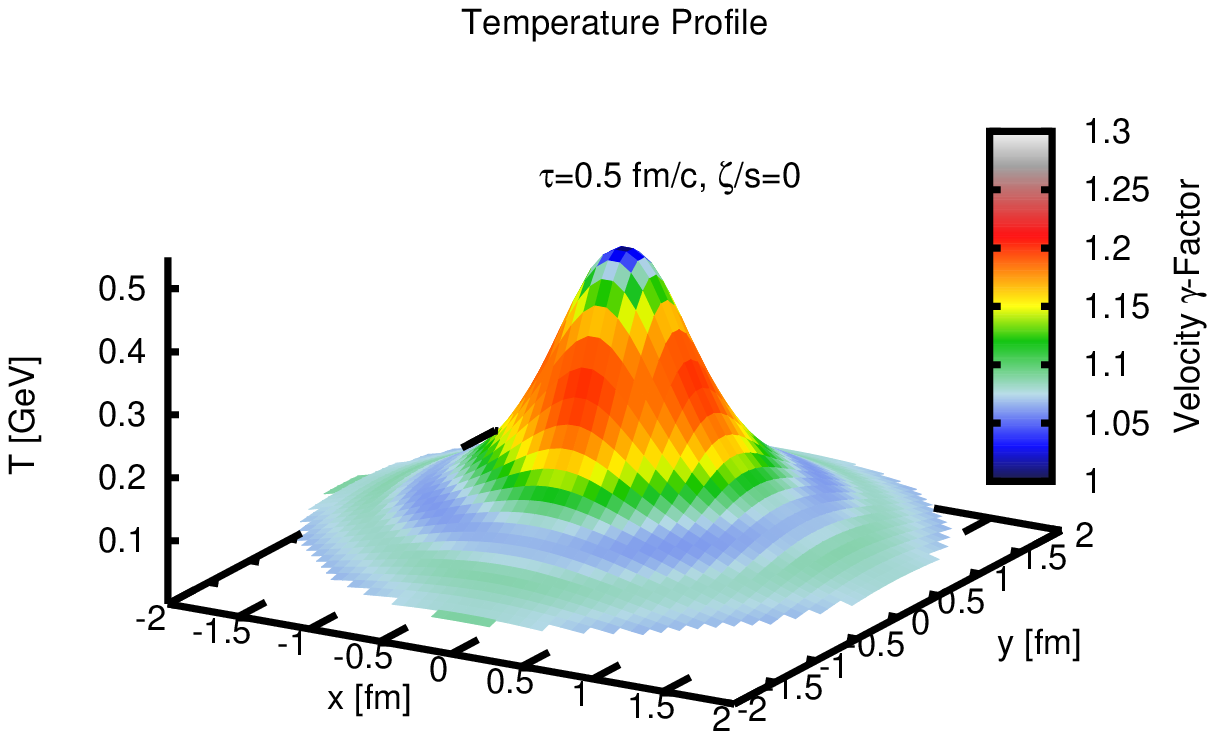}\hfill
\includegraphics[width=0.45\textwidth]{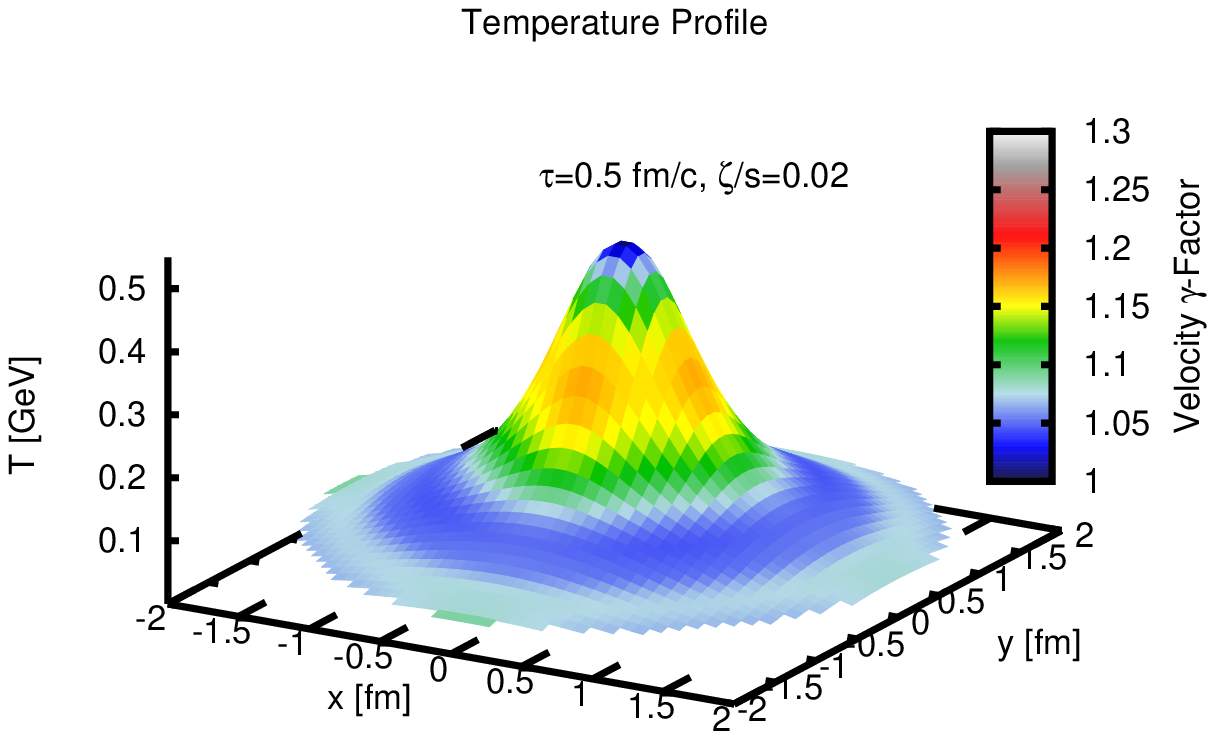}
\includegraphics[width=0.45\textwidth]{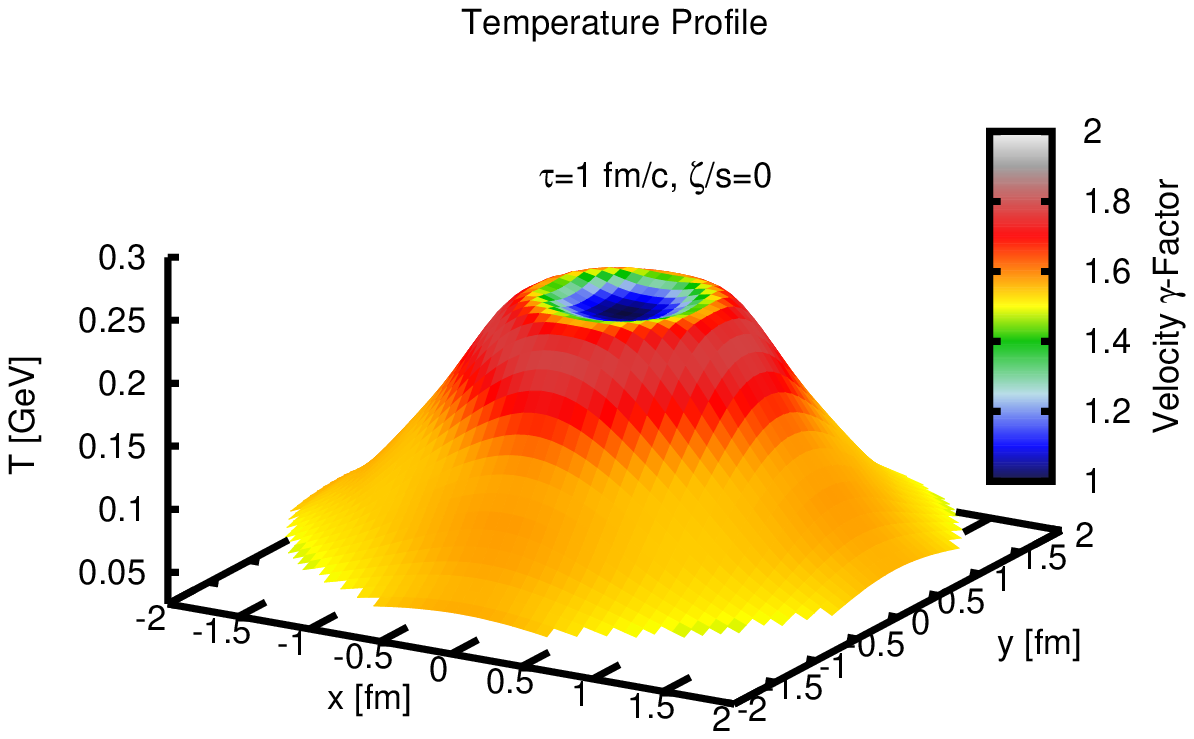}\hfill
\includegraphics[width=0.45\textwidth]{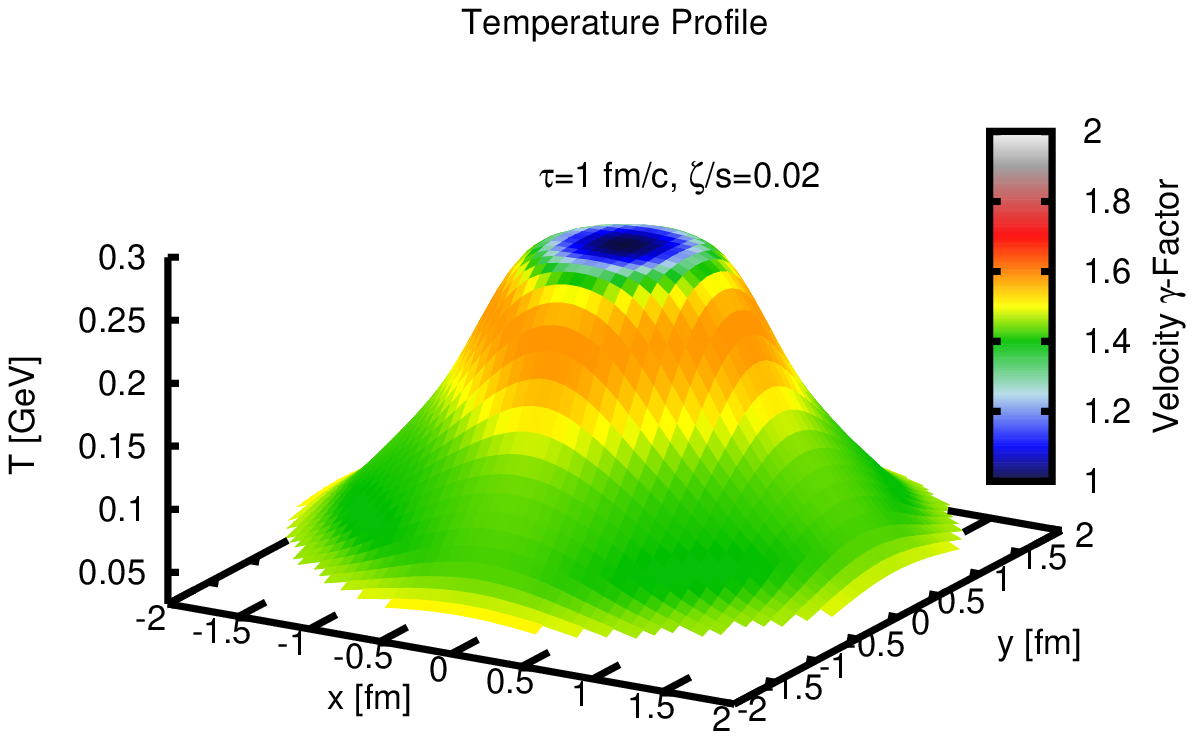}
\includegraphics[width=0.45\textwidth]{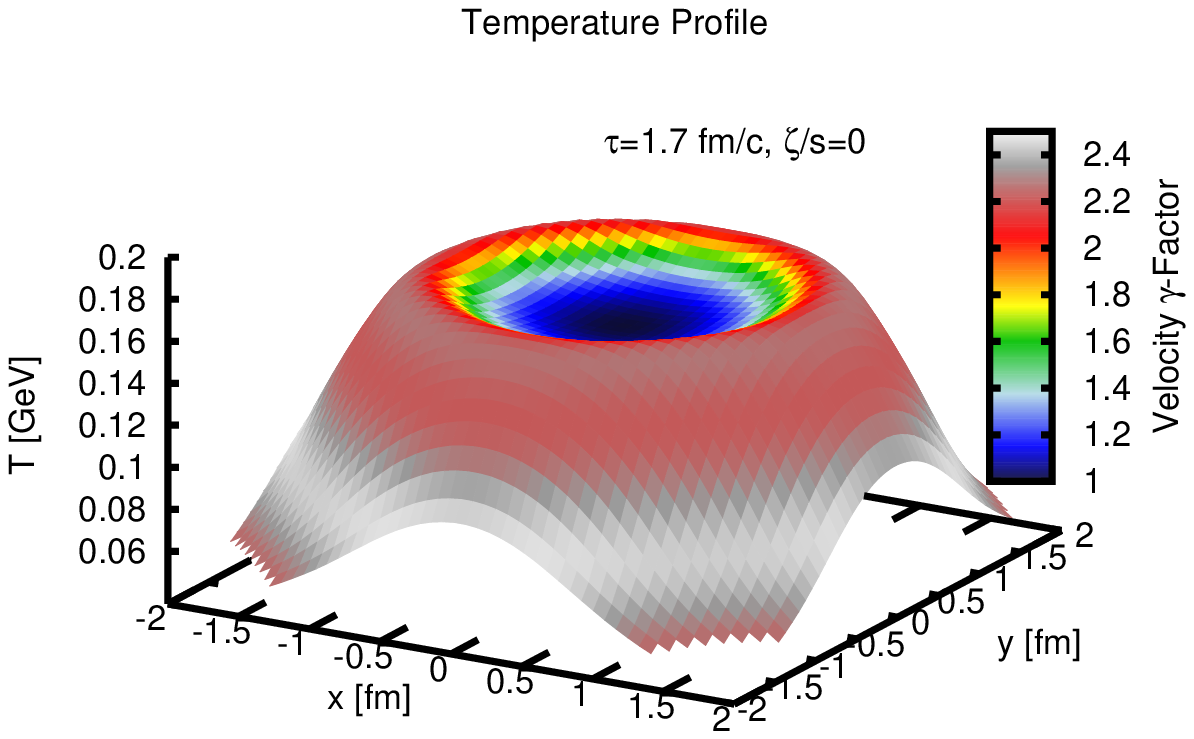}\hfill
\includegraphics[width=0.45\textwidth]{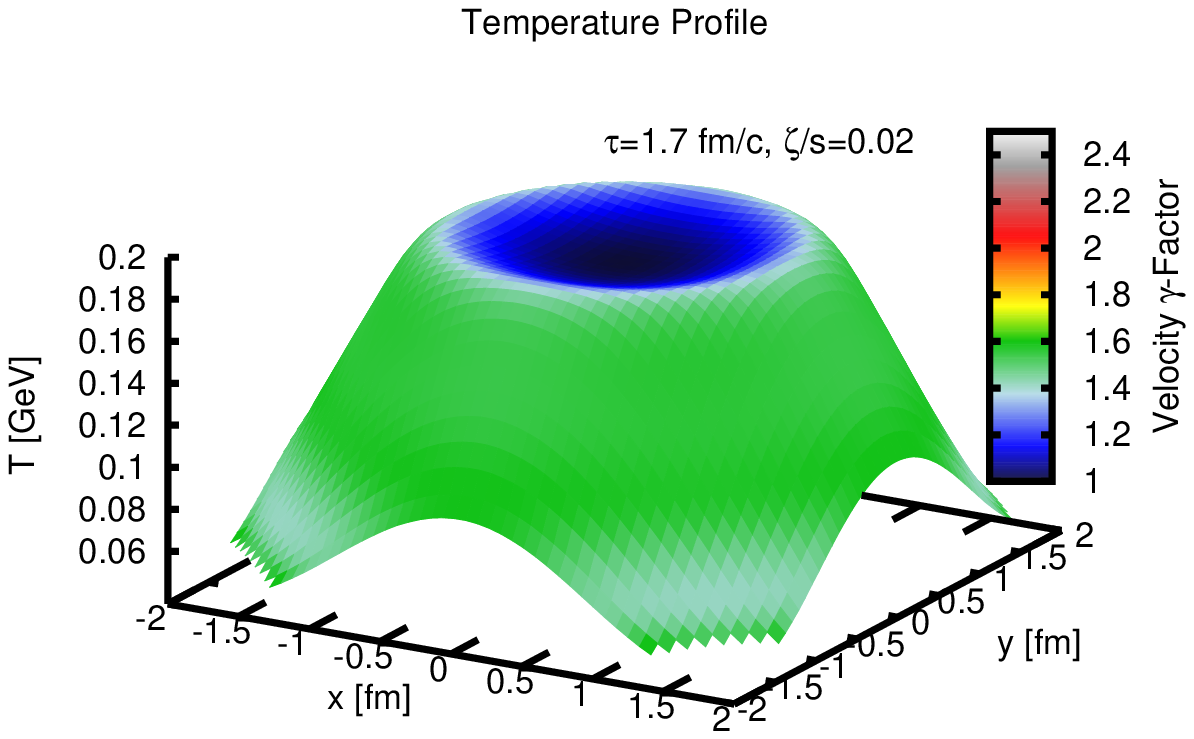}
\caption{Time-snapshot of the temperature distribution in the transverse plane, with color coding corresponding to the local fluid velocity $|{\bf v}|$ (in terms of $\gamma=\frac{1}{\sqrt{1-{\bf v}^2}}$). Left panels show results without bulk viscosity, while right panels are for $\frac{\zeta}{s}=0.02$.
\label{fig:supp}}
\end{figure*}

In the main text, it was mentioned that bulk viscosity affects the hydrodynamic 
flow pattern directly. In this appendix, the effect of bulk viscosity on the temperature and fluid velocity evolution are demonstrated through snapshots during the system evolution for an 'RND' proton collision at small impact parameter (0-10 percent centrality class), shown in Fig.~\ref{fig:supp}. The panels in the figure show that adding bulk viscosity changes the hydrodynamic evolution through reducing the local fluid velocity and slowing down the temperature decrease. Since particles are sampled from the local fluid cells, smaller velocities imply smaller particle momenta, which is consistent with the finding in the main text.

\end{appendix}
\newpage

\bibliographystyle{apsrev} 
\bibliography{flowpp}
%\newpage
%\onecolumngrid

\end{document}